\documentclass[twoside,twocolumn,english,aps,prl]{revtex4}
\usepackage[T1]{fontenc}
\usepackage[latin1]{inputenc}
\usepackage{graphicx}

\makeatletter

\usepackage{babel}
\makeatother
\begin{document}
\newcommand{\sgn}{\mathop{\mathrm{sgn}}\nolimits}

\newcommand{\const}{\mathop{\mathrm{const}}\nolimits}

\newcommand{\tr}{\mathop{\mathrm{tr}}\nolimits}

\newcommand{\ts}{\textstyle}

\newcommand{\f}[1]{\mbox {\boldmath\(#1\)}}

\title{The Intrinsic Spin Hall Conductivity in a Generalized Rashba Model}

\author{P.L. Krotkov and S. Das Sarma}

\address{Condensed Matter Theory Center, Department of Physics, University
of Maryland, College Park, MD 20742}

\begin{abstract}
We calculate the intrinsic spin Hall conductivity $\sigma^{\mathrm{sH}}$
of a two-dimensional electron system within a generalized Rashba model,
showing that it is, in general, finite and model-dependent. Considering
arbitrary band dispersion, we find that $\sigma^{\mathrm{sH}}$ in
the presence of the linear-in-momentum spin-orbit coupling of the
Rashba form does not vanish in the presence of impurities except for
the precisely parabolic spectrum. We show, using the linear response
Kubo formalism, how the exact cancellation happens for the quadratic
dispersion, and why it does not occur in general. We derive a simple
quasiclassical formula for $\sigma^{\mathrm{sH}}$ in terms of the
Fermi momenta for the two electron chiralities, and find that $\sigma^{\mathrm{sH}}$
is in general of the order of the squared strength of the Rashba term. 

\end{abstract}
\maketitle
The spin Hall effect \cite{day05} is either a brand new subject \cite{murakami03,sinova04}
or a rather old one \cite{dyakonov71} depending on one's perspective.
Much recent interest has focused on the \emph{intrinsic} spin Hall
effect (ISHE) where a non-magnetic system (e.g. a III-V semiconductor)
spontaneously exhibits a bulk spin Hall conductivity $\sigma^{\mathrm{sH}}$
(i.e., a spin conductivity transverse to the direction of an external
electric field in the absence of any applied magnetic field) arising
entirely from the spin-orbit coupling effects in the underlying one-particle
band structure. The ISHE is to be distinguished from its better-known
counterpart, the \emph{extrinsic} SHE (ESHE) \cite{engel05,tseJul05,tseAug05}
predicted by Dyakonov and Perel \cite{dyakonov71}, which arises due
to spin-correlated asymmetry in impurity scattering. Recent claims
of experimental observations of both ESHE \cite{awschalom04,sih05}
and ISHE \cite{wunderlich05} (in semiconductor structures) make the
subject particularly interesting since there is controversy about
the existence of ISHE in various theoretical models \cite{inoue04,mishchenko04,rashbaNov04,dimitrova05,raimondi05,chalaev05,rashba05,murakami05,khaetskii04,murakami04,malshukov05}.
In particular, within the extensively studied Rashba model \cite{bychkov84},
arising from the spatial inversion asymmetry in 2D systems, the intrinsic
$\sigma^{\mathrm{sH}}$ has been claimed to be a large universal constant
\cite{sinova04} $e/8\pi$, where $e$ is the electron charge, but
subsequently it was found that an arbitrarily small concentration
of impurities in the Born approximation (i.e., when one neglects all
spin-related asymmetry in scattering leading to ESHE) introduces vertex
renormalization exactly canceling \cite{inoue04,mishchenko04,dimitrova05,raimondi05,rashbaNov04}
the universal ISHE predicted in the Rashba model in \cite{sinova04}.
(The same cancellation also occurs for the 2D Dresselhaus model for
the linear-in-momentum spin-orbit coupling.) For other models of spin-orbit
coupling, e.g. the 3D Dresselhaus model \cite{malshukov05} or the
Luttinger model for the valence band holes \cite{murakami04}, the
ISHE has been theoretically calculated to be finite and model-dependent.
Since the Rashba model is by far the most widely studied model for
the spin-orbit coupling in 2D semiconductor structures and since impurities
are crucial for the validity of the linear response theory, the precise
vanishing of the ISHE in the Rashba model has led to substantial confusion
about whether the ISHE, as opposed to ESHE \cite{dyakonov71,engel05,tseJul05,tseAug05},
which is always present, is ever finite in the Rashba model \cite{inoue04,mishchenko04,dimitrova05,raimondi05,rashbaNov04}
and perhaps even in all models \cite{khaetskii04}. 

We show in this Letter that the precise vanishing of the 2D ISHE in
the Rashba model is limited to quasiclassical calculations for the
usually assumed quadratic unperturbed band spectrum \begin{equation}
\epsilon_{0}(p)=p^{2}/2m.\label{eq:quadratic}\end{equation}
 In fact, the cancellation is accidental, and slight changes in the
model (i.e., using a band dispersion different from the usual parabolic
band dispersion) give rise to a finite ISHE even in the Rashba model.
While the vertex corrections do decrease the magnitude of intrinsic
$\sigma^{\mathrm{sH}}$, they contain an integral over momenta that
depends on band dispersion $\epsilon_{0}(p)$, and $\sigma^{\mathrm{sH}}$
is finite for a general (i.e. non-parabolic) band dispersion, albeit
explicitly model-dependent. Our finding of a finite 2D ISHE in the
generalized Rashba model is of particular significance in establishing
a matter of principle, refuting the earlier findings of either the
large universal ISHE or the vanishing ISHE.

In the presence of spin-orbit coupling in a Hamiltonian, a spin current
of the from \begin{equation}
j_{i}^{\alpha}=\sigma^{\mathrm{sH}}e_{\alpha ij}E_{j}.\end{equation}
should appear in the kinetic equation for the spin density. This quantity
leads to spin accumulation at the boundary of a sample, limited by
spin diffusion, when an electric current is passed in the absence
of magnetic field (the spin Hall effect). The appearance of such a
current was predicted for a band structure of III-V semiconductors
\cite{murakami03} and for general 2D Rashba coupling \cite{sinova04},
and has been a subject of a great deal of recent activity \cite{day05}. 

We derive in this Letter the spin Hall conductivity for a general
spectrum, which has a simple quasiclassical limit. First we demonstrate
how the cancellation for the parabolic dispersion, already noted in
the literature \cite{inoue04,mishchenko04,dimitrova05,raimondi05,rashbaNov04},
takes place and also why it is not universal but explicitly depends
on the spectrum. We then provide the general quasiclassical limit
for the ISHE conductivity.

\begin{figure}
\includegraphics[%
  width=0.70\columnwidth]{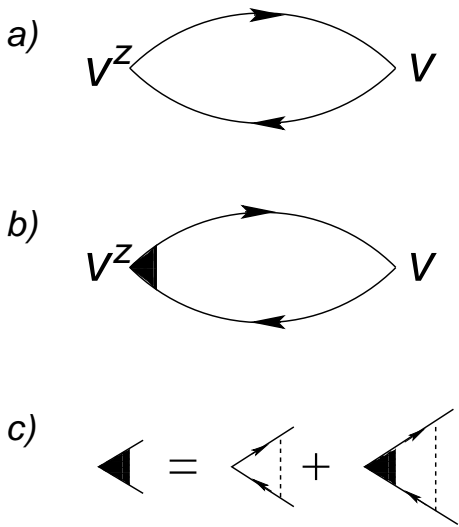}

\caption{Bubble diagrams for the spin Hall conductivity: bare (a) and with
vertex renormalized by the ladder series of impurity scattering (b).
Equation on the latter (c).\label{cap:Bubble-diagrams}}
\end{figure}

Consider the two-dimensional electron Hamiltonian\begin{equation}
\mathcal{H}=\epsilon_{0}(p)+\alpha\hat{\mathbf{z}}(\f\sigma\times\mathbf{p})\label{eq:H}\end{equation}
comprising an unperturbed part $\epsilon_{0}(p)$ and the Rashba coupling
(parameterized by the strength $\alpha$) of spin and orbital movement
in the plane defined by the normal $\hat{\mathbf{z}}$. From (\ref{eq:H})
the velocity operator is\begin{equation}
v_{i}(\mathbf{p})=\partial_{p_{i}}\mathcal{H}=\epsilon'_{0}(p)\hat{p}_{i}+\alpha e_{zji}\sigma_{j}.\label{eq:v}\end{equation}
The spin-velocity is usually defined as \begin{equation}
v_{i}^{\mu}(\mathbf{p})=\ts\frac{1}{2}\{ v_{i},\sigma_{\mu}\}=\epsilon'_{0}(p)\hat{p}_{i}\sigma_{\mu}+\alpha e_{z\mu i}.\label{eq:v^z}\end{equation}
In the framework of the linear response theory the spin Hall conductivity
can be calculated as a bubble (Fig. \ref{cap:Bubble-diagrams}) with
the vector vertices $\mathbf{j}=e\mathbf{v}$ and $\mathbf{j}^{z}=\frac{1}{2}\mathbf{v}^{z}$. 

To emphasize the cancellation we break up the spin-Hall conductivity
in two parts (the two terms in parentheses) \begin{eqnarray}
\sigma_{ij}^{\mathrm{sH}}(\Omega) & = & \frac{e}{2\Omega}T\sum_{\omega}\sum_{\mathbf{p}}\tr[v_{i}(\mathbf{p})G(\mathbf{p},\omega)\nonumber \\
 &  & \times\left(v_{j}^{z}(\mathbf{p})+V_{j}^{z}(\omega,\Omega)\right)G(\mathbf{p},\omega+\Omega)].\label{eq:sigma^sH}\end{eqnarray}
The first term is the bare bubble in Fig. \ref{cap:Bubble-diagrams}
(a), the second term is the result of the summation of the ladder
series in Fig. \ref{cap:Bubble-diagrams} (b). The terms of the ladder
series do not vanish after averaging over $\hat{\mathbf{p}}$ because
the Green's function $G(\mathbf{p},\omega)$ for the Hamiltonian (\ref{eq:H})
itself depends on the direction of momentum. 

The contribution from the first (bare) term in (\ref{eq:sigma^sH})
is $\sigma_{ij}^{\mathrm{sH}}=e_{zij}\sigma_{0}^{\mathrm{sH}}$, where
the quantity $\sigma_{0}^{\mathrm{sH}}$ is given by an integral that
depends on the spectrum $\epsilon_{0}(p)$: \begin{equation}
\sigma_{0}^{\mathrm{sH}}=\frac{e}{\Omega}T\sum_{\omega}\alpha A[\epsilon_{0}],\end{equation}
where\begin{equation}
A[\epsilon_{0}]=\frac{i}{4}\sum_{\mathbf{p}}\epsilon'_{0}(p)[G_{u+}G_{d}-G_{d+}G_{u}].\label{eq:A}\end{equation}
Here $G_{u,d}(\mathbf{p},\omega)=(i\widetilde{\omega}-\xi_{u,d}(\mathbf{p)})^{-1}$,
where $\xi_{u,d}=\epsilon_{0}(p)\mp\alpha p-\epsilon_{F}$ and $\widetilde{\omega}=\omega+\frac{1}{2\tau}\sgn\omega$,
is the Green's functions for the chiral eigenstates of the Hamiltonian
(\ref{eq:H}). For brevity we omit the arguments of the Green's function,
and denote the ones with $\omega+\Omega$ by the subscript $+$. 

Taking the integral over momenta in (\ref{eq:A}) in the $\xi$-approximation
with quadratic spectrum (\ref{eq:quadratic}) we find (see below for
details)\begin{equation}
\sigma_{0}^{\mathrm{sH}}=\frac{e}{8\pi}\frac{4\Delta^{2}\tau^{2}}{1+4\Delta^{2}\tau^{2}},\label{eq:sigma^sH_0}\end{equation}
where $2\Delta=2\alpha p_{F}$ is the value of the spin-orbital energy
gap for the two chiralities at the Fermi level. In the clean limit
$\tau\to\infty$, Eq. (\ref{eq:sigma^sH_0}) gives to the universal
constant $e/8\pi$ \cite{sinova04} independent of the strength of
the spin-orbit interaction $\alpha$. The physical meaning of this
constant can be easily seen if one carries out the frequency summation
in $\sigma_{0}^{\mathrm{sH}}$ first. Then in the clean limit one
gets \begin{equation}
\sigma_{0}^{\mathrm{sH}}(\Omega)=\frac{\alpha e}{2}\sum_{\mathbf{p}}\epsilon'_{0}(p)\frac{n_{F}(\xi_{u})-n_{F}(\xi_{d})}{(\xi_{d}-\xi_{u})^{2}+\Omega^{2}}\end{equation}
which at $T=0$ singles out the annulus $p_{d}<p<p_{u}$, populated
by the electrons with the chirality $u$ only, where $p_{u,d}=\sqrt{m^{2}\alpha^{2}+2m\epsilon_{F}}\pm m\alpha$
are the Fermi momenta for the two chiral eigenstates for the spectrum
(\ref{eq:quadratic}): \begin{equation}
\sigma_{0}^{\mathrm{sH}}(0)=\frac{\alpha e}{2}\int_{p_{d}}^{p_{u}}\frac{pdp}{2\pi}\frac{\epsilon'_{0}(p)}{(2\alpha p)^{2}}=\frac{e}{8\pi}\frac{p_{u}-p_{d}}{2m\alpha}=\frac{e}{8\pi}.\end{equation}

In (\ref{eq:sigma^sH}) the ladder summation is carried out in the
form of the renormalization of the spin-velocity vertex $\mathbf{v}^{z}\to\mathbf{v}^{z}+\mathbf{V}^{z}$.
The result for the summation of the ladder corrections is of course
independent of which of the two vertices, the current $e\mathbf{v}$
or the spin-current $\frac{1}{2}\mathbf{v}^{z}$, in (\ref{eq:sigma^sH})
is renormalized. Straightforward calculations (shown at the end of
the paper) yield the ladder correction $e_{zij}\sigma_{L}^{\mathrm{sH}}$,
where \begin{equation}
\sigma_{L}^{\mathrm{sH}}=\frac{e}{\Omega}T\sum_{\omega}A[\epsilon_{0}]\frac{\alpha B[\epsilon_{0}]-C[\epsilon_{0}]}{2\pi\nu_{0}\tau-B[\epsilon_{0}]}.\label{eq:sigma^sH_L}\end{equation}
Here \begin{eqnarray}
B[\epsilon_{0}] & = & \ts\frac{1}{4}\sum_{\mathbf{p}}(G_{u+}+G_{d+})(G_{u}+G_{d})\label{eq:B}\\
C[\epsilon_{0}] & = & \ts\frac{1}{4}\sum_{\mathbf{p}}\epsilon'_{0}(p)[G_{u+}G_{u}-G_{d+}G_{d}].\label{eq:C}\end{eqnarray}
Except for the averaging over direction $\hat{\mathbf{p}}$ of momentum,
we did not perform momentum integration in the quantities \textbf{$A[\epsilon_{0}]$},
$B[\epsilon_{0}]$ and $C[\epsilon_{0}]$. Summing up $\sigma_{0}^{\mathrm{sH}}$
and $\sigma_{L}^{\mathrm{sH}}$, we see that \begin{equation}
\sigma^{\mathrm{sH}}=\frac{e}{\Omega}T\sum_{\omega}A[\epsilon_{0}]\frac{\alpha2\pi\nu_{0}\tau-C[\epsilon_{0}]}{2\pi\nu_{0}\tau-B[\epsilon_{0}]}.\label{eq:sigma^sH_full}\end{equation}
An exact cancellation takes place \emph{only} when $C[\epsilon_{0}]=\alpha2\pi\nu_{0}\tau$.
This is the central result of the paper. Such a cancellation is indeed
what happens for the quadratic spectrum (\ref{eq:quadratic}).

In general, keeping $\epsilon_{0}(p)$ for band dispersion in the
Green's functions, we get in the $\xi$-approximation\begin{equation}
C[\epsilon_{0}]=2\pi\nu_{0}\tau\frac{I}{\Omega\tau+I}\frac{\nu_{u}\epsilon'_{0}(p_{u})-\nu_{d}\epsilon'_{0}(p_{d})}{4\nu_{0}}.\label{eq:C2}\end{equation}
Here $\nu_{u,d}$ is the density of states on the Fermi surface for
the chiralities $u$ and $d$, $\nu_{0}=(\nu_{u}+\nu_{d})/2$ by definition,
and $I=\frac{1}{2}(\sgn(\omega+\Omega)-\sgn\omega)$ is a factor accounting
for the requirement that the two poles in the Green's functions in
the bubble lie in different half-planes (i.e., one Green's function
is retarded and one advanced). If we suppose $\Omega>0$, then this
factor equals unity in the band $-\Omega<\omega<0$ and zero otherwise.
Such a factor generally appears in the frequency summation in $\sigma^{\mathrm{sH}}$,
which means that all quantities under the frequency summation should
be taken in the specified frequency band, where we may simply deem
$I=1$. For $\Omega<0$ the reasoning is similar.

We note that the density per chirality in 2D irrespective of the spectrum
is $n_{u}=p_{u}^{2}/4\pi$, hence $dn_{u}/dp_{u}=p_{u}/2\pi$. On
the other hand, using that, by definition, $\epsilon_{F}=\epsilon_{0}(p_{u})-\alpha p_{u}$,
we get $dn_{u}/dp_{u}=\nu_{u}d\epsilon_{F}/dp_{u}=$$\nu_{u}(\epsilon'_{0}(p_{u})-\alpha)$.
For the $d$ chirality, we get analogous expressions with the opposite
sign before $\alpha$. We finally get \begin{equation}
\frac{\nu_{u}\epsilon'_{0}(p_{u})-\nu_{d}\epsilon'_{0}(p_{d})}{4\nu_{0}}=\frac{\alpha}{2}\left(1+\frac{p_{u}-p_{d}}{4\pi\alpha\nu_{0}}\right).\label{eq:u_d}\end{equation}
We therefore reduced the expression that explicitly depended on the
electron spectrum by a simpler one that still depends on the spectrum
implicitly through the Fermi momenta $p_{u,d}$ for the two chiralities.
In this new form, Eq. (\ref{eq:u_d}), however, we immediately see
whether the exact cancellation for ISHE takes place for a given spectrum.
For the quadratic dispersion (\ref{eq:quadratic}) $p_{u}-p_{d}=2m\alpha=4\alpha\pi\nu_{0}$,
where $\nu_{0}=m/2\pi$, so that (\ref{eq:u_d}) gives $\alpha$,
and $C[\epsilon_{0}]=\alpha2\pi\nu_{0}\tau$ exactly cancels out the
spin Hall conductivity (\ref{eq:sigma^sH_full}).

Above we used the $\xi$-approximation in its usual form, i.e. the
prefactors before the Green's functions were taken at the Fermi level.
It may be shown that for the quadratic spectrum (\ref{eq:quadratic})
the cancellation is exact in all orders of the quasiclassical approximation,
i.e. if one expands prefactors in powers of $\xi$, all the terms
vanish.

We can calculate the total spin Hall conductivity for an arbitrary
spectrum in the quasiclassical $\xi$-approximation. Using the expressions
for $B$ and $A[\epsilon_{0}]$ obtained in the end of the paper,
we get for the sum of the bare and ladder contributions:\begin{equation}
\sigma_{ij}^{\mathrm{sH}}=e_{zij}\frac{e}{8\pi}\left(\frac{p_{u}-p_{d}}{4\pi\alpha\nu_{0}}-1\right).\label{eq:quasiclassical}\end{equation}

For a spectrum different from (\ref{eq:quadratic}), the cancellation
in the expression above does not take place, e.g., if $\epsilon_{0}(p)=v_{0}p,$
then $p_{u,d}=\epsilon_{F}/(v_{0}\pm\alpha)$ and \begin{equation}
\nu_{0}=\frac{\nu_{u}+\nu_{d}}{2}=\frac{\epsilon_{F}}{2\pi}\frac{v_{0}^{2}+\alpha^{2}}{v_{0}^{2}-\alpha^{2}}.\end{equation}
So that $p_{u}-p_{d}=4\pi\nu_{0}\alpha(v_{0}^{2}-\alpha^{2})/(v_{0}^{2}+\alpha^{2})$,
whence $\sigma^{\mathrm{sH}}=-(e/8\pi)2\alpha^{2}/(v_{0}^{2}+\alpha^{2})$.

It may be shown that the net $\sigma^{\mathrm{sH}}$ is always at
most of the order $\alpha^{2}$. Expanding $p_{u,d}$ in (\ref{eq:quasiclassical})
in powers of $\alpha$ around the Fermi momentum $p_{F}$ of the general
band dispersion $\epsilon_{0}(p_{F})=\epsilon_{F}$, we find that
the lowest order expression always cancels the unity in the parentheses
of Eq. (\ref{eq:quasiclassical}), and leaves the next order term,
$O(\alpha^{2})$, as the leading ISHE contribution for small $\alpha$.

Below we provide details of the derivation of our results above. The
Hamiltonian (\ref{eq:H}) can be diagonalized:\begin{equation}
U^{+}\mathcal{H}U=\epsilon_{0}(p)-\alpha p\sigma_{z},\label{eq:RotdH}\end{equation}
 by rotation in spin space: \begin{equation}
U=\frac{1}{\sqrt{2}}\left(\begin{array}{cc}
1 & 1\\
ie^{i\varphi_{\hat{\mathbf{p}}}} & -ie^{i\varphi_{\hat{\mathbf{p}}}}\end{array}\right)\end{equation}
where $\varphi_{\hat{\mathbf{p}}}$ is the angle between $\hat{\mathbf{p}}$
(the direction of electron momentum) and $\hat{\mathbf{x}}$. The
eigenstates of the rotated Hamiltonian (\ref{eq:RotdH}) are spin
up and spin down. In the original basis they correspond to the two
chiralities (spin along and opposite to $\hat{\mathbf{z}}\times\hat{\mathbf{p}}$).
Columns of $U$ give the eigenfunctions in the original basis.

In the basis of the chiral eigenstates the Green's function, which
is a matrix in spin space, is diagonal with the elements $G_{u,d}$.
In the original frame \begin{eqnarray}
G(\mathbf{p},\omega) & = & U\left(\begin{array}{cc}
G_{u} & 0\\
0 & G_{d}\end{array}\right)U^{+}\nonumber \\
 & = & \ts\frac{1}{2}(G_{u}+G_{d})-\frac{1}{2}(G_{u}-G_{d})e_{zij}\sigma_{i}\hat{p}_{j}.\label{eq:Green's}\end{eqnarray}

The renormalized vertex $\mathbf{V}^{z}$ is found from an inhomogeneous
linear equation 

\begin{equation}
\mathbf{V}^{z}=\frac{1}{2\pi\nu_{0}\tau}\sum_{\mathbf{p}}G(\mathbf{p},\omega)\left(\mathbf{v}^{z}(\mathbf{p})+\mathbf{V}^{z}\right)G(\mathbf{p},\omega+\Omega),\label{eq:V^z}\end{equation}
represented by the diagrammatic equation in Fig. \ref{cap:Bubble-diagrams}
(c). Here $1/2\pi\nu_{0}\tau$ is the impurity line in the Born approximation.
The inhomogeneous term, being a spin matrix, has only off-diagonal
elements ($i=x,y$)

\begin{equation}
\sum_{\mathbf{p}}G(\mathbf{p},\omega)v_{i}^{z}(\mathbf{p})G(\mathbf{p},\omega+\Omega)=\sigma_{i}A[\epsilon_{0}].\end{equation}
Hence $\mathbf{V}^{z}$ only has the same components (that depend
on $\omega$, $\Omega$ and not on \textbf{$\mathbf{p}$}):\begin{equation}
\mathbf{V}^{z}=\frac{\f\sigma A[\epsilon_{0}]}{2\pi v_{0}\tau-B}.\end{equation}
Calculating the ladder contribution to the conductivity with the renormalized
vertex $\mathbf{V}^{z}$ we get (\ref{eq:sigma^sH_L}).

Finally, we calculate in the $\xi$-approximation the quantities $A[\epsilon_{0}]$,
$B[\epsilon_{0}]$ and $C[\epsilon_{0}]$ introduced in (\ref{eq:A}),
(\ref{eq:B}) and (\ref{eq:C}). The $\xi$-approximation consists
in replacing $\sum_{\mathbf{p}}\to\nu\int d\xi$, all prefactors before
the Green's functions being taken at the Fermi level. The pairs of
Green's functions for the same chirality are the easiest:\begin{equation}
\sum_{\mathbf{p}}G_{u+}G_{u}=\int\frac{\nu_{u}d\xi_{u}}{(\xi_{u}-i\widetilde{(\omega+\Omega)})(\xi_{u}-i\widetilde{\omega})}=\frac{2\pi\nu_{u}\tau I}{\Omega\tau+I},\end{equation}
and analogously for the $d$ chirality. The cross-products on the
other hand depend on the gap $\Delta$ at the Fermi surface $\xi_{u,d}=\xi\mp\Delta$:\begin{equation}
\sum_{\mathbf{p}}G_{u+}G_{d}=\frac{2\pi\nu_{0}\tau I}{\Omega\tau+I-i2\Delta\tau},\end{equation}
and the expression with $\Delta\leftrightarrow-\Delta$ for the crossproduct
$G_{d+}G_{u}$. Hence we find, taking $\Omega\tau\to0$ and in the
band of frequencies where $I=1$: \begin{eqnarray}
A[\epsilon_{0}] & = & -2\pi\nu_{0}\epsilon'_{0}(p_{F})\frac{\Delta\tau^{2}}{1+4\Delta^{2}\tau^{2}},\label{eq:A2}\\
2\pi\nu_{0}\tau-B[\epsilon_{0}] & = & 2\pi\nu_{0}\tau\frac{2\Delta^{2}\tau^{2}}{1+4\Delta^{2}\tau^{2}},\label{eq:B2}\end{eqnarray}
and the expression (\ref{eq:C2}) for $C[\epsilon_{0}]$. The product
$2\pi\nu_{0}\epsilon'_{0}(p_{F})=p_{F}$, so that $\alpha A[\epsilon_{0}]=-\Delta^{2}\tau^{2}/(1+4\Delta^{2}\tau^{2})$. 

In summary, we have shown that the vanishing of the 2D ISHE within
the Rashba model of spin-orbit coupling is specific to both the linear-in-momentum
Rashba coupling \cite{inoue04} and to the parabolic energy band dispersion.
For a general band dispersion, the ISHE within the Rashba model is
finite, albeit strongly dependent not only on the Rashba coupling
$\alpha$ but also on the details of the band dispersion. In general,
the impurity vertex corrections suppress the intrinsic $\sigma^{\mathrm{sH}}$
from its universal value of $e/8\pi$, reducing it to some small nonuniversal
finite value of order $\alpha^{2}$. Our result, combined with recent
theoretical demonstrations of finite ISHE in the 3D Dresselhaus model
\cite{malshukov05} and the Luttinger model \cite{murakami04}, establish
that the intrinsic spin Hall effect is theoretically well-defined,
but non-universal.

This work is supported by NSF, US-ONR, and LPS-NSA.

\end{document}